\documentclass []{article}
\usepackage{amsmath}
\usepackage{epsfig}
\usepackage{amsfonts}

\begin{document}

\title{Emergence of the Dirac equation in the solitonic\\ 
source of the Kerr spinning particle}

\author{A. Burinskii \\
    Lab. of Theor. Phys., NSI Russian Academy of
Sciences,\\ B. Tulskaya 52  Moscow, 115191 Russia, bur@ibrae.ac.ru}

\maketitle

\begin{abstract}Gravitational and electromagnetic (EM)
    field of the Dirac electron is described by the  Kerr-Newman (KN) solution.
    We elaborate a regular source of the KN solution which satisfies the requirement
    of flat space-time inside the source and realization of the exact KN solution
    outside the source. This requirement removes conflict between gravity and quantum theory and determines many details of the source structure. In particular, we obtain that
    the KN source should forms a gravitating bag model, similar  to the known MIT and SLAC bag models. As opposite to the known bag models, the self-interacting Higgs field should be confined inside the bag, while outside the bag the gauge symmetry should be unbroken to provide the external KN gravity.
    We show that twistorial structure of the Kerr geometry (the Kerr theorem) determines
    structure of  the Dirac equation, resulting in a variable mass term, which is generated inside the bag
    through interaction with the confined Higgs field. Similar to the other bag models,
    ellipsoidal deformations of the KN bag creates a string-like structure of the dressed electron
    -- circular string  positioned along perimeter of the KN bag.
\end{abstract}

\def\b{\bar}
\def\d{\partial}
\def\D{\Delta}
\def\cD{{\cal D}}
\def\cK{{\cal K}}
\def\f{\varphi}
\def\g{\gamma}
\def\G{\Gamma}
\def\l{\lambda}
\def\L{\Lambda}
\def\M{{\Cal M}}
\def\m{\mu}
\def\n{\nu}
\def\p{\psi}
\def\q{\b q}
\def\r{\rho}
\def\t{\tau}
\def\x{\phi}
\def\X{\~\xi}
\def\~{\widetilde}
\def\h{\eta}
\def\bZ{\bar Z}
\def\cY{\bar Y}
\def\bY3{\bar Y_{,3}}
\def\Y3{Y_{,3}}
\def\z{\zeta}
\def\Z{{\b\zeta}}
\def\Y{{\bar Y}}
\def\cZ{{\bar Z}}
\def\`{\dot}
\def\be{\begin{equation}}
\def\ee{\end{equation}}
\def\bea{\begin{eqnarray}}
\def\eea{\end{eqnarray}}
\def\half{\frac{1}{2}}
\def\fn{\footnote}
\def\bh{black hole \ }
\def\cL{{\cal L}}
\def\cH{{\cal H}}
\def\cF{{\cal F}}
\def\cP{{\cal P}}
\def\cM{{\cal M}}
\def\ik{ik}
\def\mn{{\mu\nu}}
\def\a{\alpha}

{
\def\oto{\overrightarrow}
\def\cA{{\mathcal A}}


\section{Introduction} It has been discussed for long time that
black holes (BH) have to
be related with elementary particles \cite{BHpart}. However, spin
and charge of particles prevent formation of the BH horizons. A BH
looses the horizons if the charge $e$ or spin parameter $a=J/m$
exceeds the mass $m$ (in the dimensionless units $G=c=\hbar =1 $).
For example, the electron charge exceeds the mass for 21 order,
while its spin/mass ratio is about $10^{22}$, and the BH threshold
$a \le m $ is exceeded for $44$ orders. Similar relations are
valid for the other elementary particles, and besides the Higgs
boson, which has neither spin nor charge, none of the elementary
particles may be associated with a black hole. Meanwhile, it does
not means that it concerns the over-rotating BH geometry without
horizons.

 As it was shown by Carter \cite{Car}, the Kerr-Newman rotating
BH solution has gyromagnetic ratio $g=2$ as that of the Dirac
electron, and  the four measurable parameters of the electron:
spin, mass, charge and magnetic moment shows unambiguously  that
 gravitational and electromagnetic field of the electron should
correspond to over-rotating Kerr-Newman (KN) solution. The
corresponding space has topological defect -- the naked Kerr
singular ring, which forms a branch line of space
 into two sheets: the sheet of advanced and sheet of the retarded
fields. The Kerr-Schild form of metric \be g_\mn =\eta_\mn + 2H
k_\m k_\n ,\label{KSH} \ee in which $ \eta_\mn $ is metric of
auxiliary Minkowski space $M^4 ,$ and $ k_\m $ is a null vector
field, $ k_\m k^\m =0 ,$ forming the Principal Null Congruence
(PNC) $\cal K .$\fn{We use signature $(- + + +)$.} These retarded
and advanced sheets are related by analytic transfer of the PNC
via disk $ r=0 $ spanned by the Kerr singular ring $ r=0, \
\cos\theta=0 $ (see fig.1). So far as $r$ is the Kerr ellipsoidal
radial coordinate,  the surface $r=0$ represents a disklike "door"
from negative sheet $r<0$ to positive one $r>0$. The null vector
fields $k^{\m\pm}(x)$ differ on these sheets, and form the
different null congruences ${\cal K}^\pm ,$ creating different
metrics \be g_\mn^\pm =\eta_\mn + 2H k_\m^\pm k_\n^\pm
\label{KSpm} \ee on the same Minkowski background $M^4.$
  This mysterious twosheetedness  caused search for different models
  of the source of Kerr geometry without negative sheet.

\begin{figure}[tp]
\centering
\includegraphics[width=3.2in]{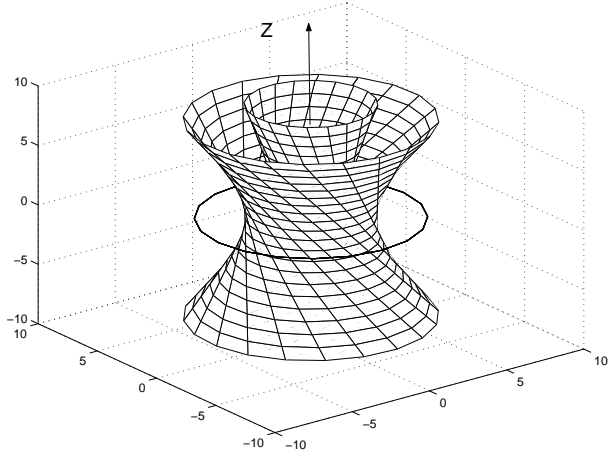}
\caption{\small Kerr's principal congruence of the null lines
(twistors) is focused on the Kerr singular ring, forming a branch
line of the Kerr space into two sheets.} \label{Liklpict}
\end{figure}

Singular metric conflicts with basic principles of quantum theory
which is settled on the flat space-time and negligible
gravitation. Resolution of this conflict requires "regularization"
of space-time, which has to be done \emph{before quantization,
i.e. on the classical level}. Singular region has to be excised
and replaced by a regular core with a flat internal metric
$\eta_\mn ,$ matching with external KN solution. Long-term search
for the models of regular source (H. Keres (1966), W. Israel
(I970), V. Hamity (1976), C. L\'opez (1984) at al.)
\cite{Keres,Isr,Ham,Lop} resulted in appearance of the gravitating
soliton model \cite{BurSol} which represents a domain-wall bubble, or a bag
 confining the Higgs field in a superconducting false-vacuum
state. Such a matter regulates the KN electromagnetic (EM) field
pushing it from interior of the bag to domain wall boundary and
results in the consistency with  flat internal metric required by
Quantum theory. The Higgs mechanism of broken symmetry approaches
this model with the known models of MIT- and SLAC- bags, and with
 the Coleman Q-ball models  \cite{GRos,Coleman}, considered as electroweak soliton models in
\cite{Kusen,VolkWohn,Grah}. The used in MIT- and SLAC- bag models
quartic potential for the self-interacting Higgs field $\Phi$,
\be V(|\Phi|)=g(\bar\sigma \sigma - \eta^2)^2 ,\label{phi4} \ee
describes a spontaneously broken theory, in which
 vacuum expectation value (vev) of the Higgs field $\sigma =<|\Phi|>$
 vanishes inside the bag, $r<R , $ and takes nonvanishing value $ \sigma = \eta ,$
 \emph{ outside the bag,} $r> R .$
The Dirac equation of the SLAC -bag theory in the presence of the
classical $\sigma$-field takes the form \be (i\gamma ^\m \d_\m -
g\sigma) \psi =0 \label{Dir-sigma}, \ee where $g$ is a dimensionless coupling parameter.
This expression shows that the
Dirac field $\psi$ acquires effective mass $m =g\sigma $ from
the vev of Higgs field $\sigma $. Inside the bag the Dirac field
is massless, while outside the bag the wave function $\psi$ may
acquire large mass $m=g\eta .$ The quarks are confined, preferring
a more  favorable energetic position inside the bag, which is the
principal idea of the confinement mechanism.

\section{Supersymmetric phase transition and formation of  false-vacuum bubble.} Such a structure of the broken symmetry is not appropriate for
the gravitating KN soliton model, since the vev of Higgs field
$\sigma$ breaks also the gauge symmetry the gravitational and
electromagnetic (EM) external KN fields, turning them into
short-range ones. An opposite (dual) geometry is realized in the
Coleman's Q-ball models \cite{GRos,Coleman}, in which the Higgs
field is confined inside the ball, $ r<R ,$ and the external
vacuum state is unbroken. However, formation of the corresponding
potential turns out to be a very non-trivial problem, and  we have
showed in \cite{BurSol} that this type of broken symmetry may be
obtained by using a supersymmetric scheme of phase transition
containing the three chiral fields $\Phi^{(i)}, \ i=1,2,3 ,$
\cite{WesBag}. One of this fields, say $\Phi^{(1)} ,$ has the
required radial dependence, and we chose it as the Higgs field
${\cal H} ,$ setting the additional notations in accord with \be
({\cal H}, Z, \Sigma) \equiv (\Phi^0, \Phi^1, \Phi^2) . \ee
 The required potential \be V(r)=\sum _i |\d_i W|^2  \ee is obtained
 from the superpotential (suggested by J.Morris in \cite{Mor})
 \be W(\Phi^i, \bar \Phi^i) = \lambda Z(\Sigma \bar \Sigma -\eta^2)
+ (Z+ \m) {\cal H} \bar {\cal H},\label{W} \ee where $ \m, \ \eta,
\ \lambda $ are real constants.
 The condition \be \d_i W =0 \ee determines two vacuum states
 separated by a spike of the potential $V$ at $r=R$:

(I) external vacuum, $r>R $, $V (r) = 0 ,$ with vanishing Higgs field ${\cal H} =0 $,   and

(II) internal  vacuum state, $r<R $, $V (r) = 0 ,$ which is indeed
a false vacuum state, since the Higgs field is not vanish, $|{\cal
H}| = \eta \lambda ^{-1/2} = const. ,$ and therefore symmetry is
broken.

Domain wall boundary of the phase transition between the states (I) and (II)  is determined by matching the external KN metric
 $ g_\mn =\eta_\mn + 2H k_\m k_\n , $ where \be
H=\frac {mr -e^2/2}{r^2+a^2 \cos ^2 \theta} \ee with flat internal
metric $ g_\mn =\eta_\mn .$ It fixes the boundary at $H=0 ,$ or
 $ r=R = \frac {e^2}{2m} .$
 Since $r$ is the Kerr oblate coordinate,
 the bag forms an oblate disk of the radius
 $ r_c \approx a = \frac {1}{2m}$ with thickness $r_e=\frac {e^2}{2m},$
so that  $r_e/r_c = e^2 \approx 137^{-1}.$

\section{Advanced fields and two sheets outside the KN source.}
The KN solution may be represented in the Kerr-Schild (KS) form via the both Kerr congruences $k^+_\m $ or
 $k^-_\m ,$ but not via the both ones simultaneously, \cite{MTW,BurA}. Vector potential $A_\m$ of the KN solution is also to be aligned with the Kerr congruence, and by the use of $k^+_\m $ or $k^-_\m $ congruence, it turns out to be either retarded, $A_{ret}$, or advanced,  $A_{adv} .$ For the \emph{physical sheet} of the
 KN solution we chose the outgoing Kerr congruence $k^+_\m ,$ corresponding to the retarded EM field $A_{ret} .$
  The fields $A_{ret}$ and $A_{adv}$ cannot reside on the same physical sheet, because each of them should be aligned with the corresponding Kerr congruence. Considering the retarded sheet as a basic physical
sheet, we fix the congruence $k^+_\m $ and the corresponding metric $g_\mn^+$, which are not allowed for the advanced field $A_{adv}.$ The field $A_{adv}$ is to be compatible with another congruence $k^-_\m ,$  positioned on the separate sheet which different metric  $g_\mn^-$.
 It should be emphasized, that this problem disappears inside the bag, where $H=0, $ and the space is
flat, $g^\pm = \eta_\mn ,$   and the difference between two metrics  disappears. Therefore, the regulated KN space-time takes the twosheeted structure outside the bag, as it is illustrated on Fig.2.

\begin{figure}[tp]
\centering
\includegraphics[width=3.2in]{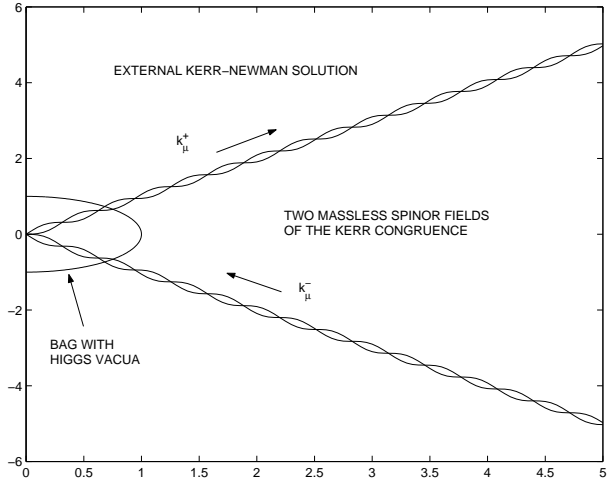}
\caption{\small Two sheets of external KN solution are matched
with flat space inside the bag. The massless spinor fields
$\phi_\alpha$ and $\bar\chi^{\dot\alpha} $ live on different KN
sheets, aligned with $k^+_\m$ and $k^-_\m $ null directions.
Inside the bag they
 join into Dirac bispinor,  getting mass from the Yukawa coupling.  } \label{Liklpict}
\end{figure}

We obtain that removing the twosheetedness related with the source of KN solution, we meet it
again from another side related with advanced potentials outside the regulated bag-like source of KN solution.
\fn{Following Dirac and Feynman \cite{DirRad,FeynST}, the
retarded potentials $A_{ret}$ can be split into a half-sum and half-difference
with advanced ones $A_{adv}$  $ A_{ret} = \frac 12
[A_{ret} + A_{adv}] + \frac 12 [A_{ret} -
A_{adv}],$ with setting correspondence of the half-difference
with radiation reaction and the half-sum with a self-interaction of the source.}
We discuss here this new effect in details, because it turns out to be related with solutions of
the Dirac equation on the KS background.

The Kerr congruences are determined by \emph{the Kerr theorem},
\cite{DKS,KraSte,PenRin,BurTMP}, which presents for the KN
solution two different congruences $k_\m^\pm ,$
\cite{BurMag,BurTMP}. The considered in sec.1 twosheeted structure
of the source was related with one of the congruences, $k_\m^+ .$
The second sheet of metric was created by \emph{analytic extension
of this congruence} to negative sheet of the KN solution
corresponding to $r<0 .$  The considered now twosheeted structure
has another origin. The two congruences $k^\pm_\m$ are now related
with \emph{two different solutions of the Kerr theorem}.

 \section{The Kerr theorem}
 Kerr theorem determines all the geodesic and \emph{shear free}
 congruences as analytical solutions of the equation
\be F(T^A) =0 ,\label{FTA}\ee where $F$ is an arbitrary
holomorphic function of the projective twistor variables \be T^A=
\{ Y, \ \z - Y v, \ u + Y \Z \}, \qquad A=1,2,3 , \label{(TA)} \ee
where $\z = (x+iy)/\sqrt 2, \ \Z = (x-iy)/\sqrt 2 , \ u = (z +
t)/\sqrt 2 , \ v = (z - t)/\sqrt 2 $ are  null Cartesian
coordinates of the auxiliary Minkowski space.

We notice, that the first twistor coordinate $Y$ is also a
projective spinor coordinate \be Y =\phi_1/\phi_0 , \label{Y10}
\ee and it is equivalent to two-component Weyl spinor $\phi_\alpha
,$ which defines the null direction\fn{We use the spinor notations
of the book \cite{WesBag}, where the $\sigma$-matrixes has the form
$\sigma^\m =(1, \sigma^i), \ \bar\sigma^\m =(1,  - \sigma^i), \ i=1,2,3 $ and
$\sigma^\m =\sigma^\m_{\alpha \dot \alpha} ,\ \bar\sigma^\m
=\bar\sigma^{\m \dot\alpha \alpha} $.}
   $ k_\m = \bar
\phi_{\dot\alpha} \sigma_\m^{\dot\alpha \alpha}\phi_\alpha .$

It is known, \cite{DKS,BurMag,BurTMP}, that function $F$ for the
Kerr and KN solutions may be represented in the quadratic in $Y$
form, \be F(Y,x^\m) = A(x^\m) Y^2 + B(x^\m) Y + C(x^\m).
\label{FKN} \ee In this case  (\ref{FTA}) can explicitly be
solved, leading to two solutions
 \be Y^\pm (x^\m)= (- B \mp \tilde r )/2A, \label{Ypm}\ee
where $\tilde r= (B^2 - 4AC)^{1/2} .$ It has been shown in \cite{BurTMP}, that these solutions are antipodally
conjugate, \be Y^+ = - 1 /{\bar Y^-} \label{antipY}.
\ee

Therefore, the solutions (\ref{Ypm}) determine two Weyl spinor
fields $\phi_\alpha $  and $\bar\chi_{\dot\alpha}$, which in
agreement with (\ref{antipY}) are related with  two antipodal
congruences \be Y^+ = \phi_{1}/\phi_{0} \label{Y+10} ,\ee \be Y^-
= \bar\chi_{\dot 1}/\bar\chi_{\dot 0} \label{Y-10} .\ee
 In the Debney-Kerr-Schild (DKS) formalism \cite{DKS} function $Y$ is
also a \emph{projective angular coordinate} $Y^+ = e^{i\phi} \tan
\frac \theta 2 .$ It gives to spinor fields $\phi_\alpha$ and
$\bar\chi_{\dot\alpha}$ an
 explicit  dependence on the Kerr angular coordinates $\phi$ and $\theta .$

For the congruence $Y^+ $ this dependence takes the form \be \phi_{\alpha} =
\left(\begin{array}{c}
e^{i\phi/2} \sin \frac \theta 2   \\
e^{-i\phi/2} \cos \frac \theta 2
\end{array} \right) . \label{phiY+}\ee
In agreement with (\ref{antipY}) we have $\Y^- = - e^{-i\phi} \cot
\frac \theta 2 ,$ and from the Lorentz invariant normalization
$\phi_\alpha \chi^\alpha = 1$ we obtain $ \chi_{\alpha} =
\left(\begin{array}{c}
 - e^{i\phi/2} \cos \frac \theta 2   \\
 e^{-i\phi/2} \sin \frac \theta 2
\end{array} \right)$ which yields \be \bar\chi^{\dot\alpha}
= \epsilon ^{\dot\alpha \dot\beta} \bar\chi_{\dot\beta} =
\left(\begin{array}{c}
 e^{i\phi/2} \sin \frac \theta 2   \\
 e^{-i\phi/2} \cos \frac \theta 2
\end{array} \right) \label{phiY-} .\ee

These massless spinor fields can be connected to the left-handed
and right-handed congruence, and only one of them, say ``left'',
$k^{(+)}_\m(x) $ is ``retarded'' and corresponds to the external
KN solution. In DKS formalism, the vector field $k^{(\pm)}_\m(x) $
is determined  by the differential form \be k_\m dx^\m =P^{-1} (du
+ \bar Y d \zeta + Y d \bar\zeta - Y\bar Y dv) , \label{e3} \ee
where $ P=(1+Y\Y)/\sqrt 2$ may be considered as a normalizing
factor for the time-like component , $k^{(\pm)}_0(x) = 1 .$
Antipodal map (\ref{antipY}) transforms the normalized field
$k^{(+)}_\m(x)= (1, \textbf{k}) $ in the field $k^{(-)}_\m(x)= (1,
- \textbf{k}) ,$ which retains the time-like direction and
reflects the space orientation. Therefore, the spinor fields
created by the Kerr theorem
 $\phi_\alpha$ and $\bar\chi^{\dot\alpha}$ correspond to the left
 out-field and right-in fields, i.e. to the retarded and advanced
 fields correspondingly.

\section{Dirac equation and two solutions of the Kerr theorem.}
The KN solution belongs to the class of algebraically
special Kerr-Schild (KS) solutions, for which all the tensor
quantities are to be aligned with  null directions of the Kerr
congruence $k_\m .$ In means that the consistent solutions of the
Dirac equation on the KS background should be aligned with the
Kerr congruence. It has been showed in  \cite{EinFinkel}  that the
Dirac field aligned with KS background should satisfy the
linearized Dirac equations \be
 \sigma ^\m _{\alpha \dot \alpha} i \d_\m
 \bar\chi ^{\dot \alpha}=  m \phi _\alpha , \quad
 \bar\sigma ^{\m \dot\alpha \alpha} i \d_\m
 \phi _{\alpha} =  m \bar\chi ^{\dot \alpha},
\label{Dir} \ee in which gravity drops out. For the Dirac bispinor
$\Psi = \left(\begin{array}{c}
 \phi _\alpha \\
\bar\chi ^{\dot \alpha}
\end{array} \right),$ the alignment
conditions  $k^{\m} \gamma_\m \Psi =0  $ turn into
equations for eigenfunctions of the helicity operator
$(\mathbf{k}\cdot \mathbf{\sigma})$ \cite{BLP}, \be
(\mathbf{k}\cdot
 \mathbf{\sigma})  \phi = \phi, \quad (\mathbf{k}\cdot \mathbf{\sigma})  \chi = - \chi,
\label{align} \ee and one sees that the spinor fields $\phi$ and
$\chi$ have opposite helicity, forming the "left-handed"  $\phi$
and "right-handed" helicity states, aligned with out-going
direction $\mathbf{k}$ and in-going direction $-\mathbf{k}$
correspondingly. In Kerr geometry, these fields should be placed
on different sheets corresponding to two antipodal congruences
$k_\m^\pm $ obtained from the Kerr theorem.  Authors of the paper
\cite{EinFinkel} concluded that these solutions \emph{``are not
consistent unless the mass vanishes..."}. Indeed, the left-handed
part of the Dirac equation is aligned with physical sheet of the
KN geometry, while the right-handed parts is aligned with the
second sheet obtained under parity inversion of the Kerr null
congruence. For the zero mass, the left- and right-hand parts of
the Dirac equations decouple, leading to solutions with opposite
helicity which are consistent with different sheets of the KN
geometry. In the same time,  the both null congruences $k^\pm_\m$
coexist without conflict \emph{on the flat space-time}, where
 the massive Dirac equation is consistent with the
both Kerr congruences.

In particular,  there exist in flat space-time the massive plane
wave solutions \cite{BLP} (v.1, sec. 16 and sec. 23), identified
as the \emph{spherical helicity states} \be \Psi_p= \frac 1
{\sqrt{2\epsilon}} u_p \exp^{-ipx} , \label{PsiPlane}\ee where
$\epsilon= + \sqrt{p^2 +m^2} $,   $p$ is 4-momentum and $u_p$ is
the normalized bispinor formed from (\ref{phiY+}) and
(\ref{phiY-}).

Therefore, the massive Dirac solutions aligned with the both Kerr
null directions exist only inside the bag, where the spice-time is
flat. Outside the bag, the KN gravitational field breaks parity of
the left- and right-handed spinors, and the Dirac bispinor splits
into the massless left- and right- Weyl spinors which should be
placed on the different sheets of the KN solution, as it is
illustrated in Fig.2.

\section{Variable mass and the bag model conception}
We arrive at the Dirac equation with a variable mass
term which changes for different regions of the space-time. We
notice that it is a proper feature of the MIT- and SLAC- bag
models related with principal idea of the quark confinement
\cite{MIT,SLAC}. The quark wave function, solution of the Dirac
equation with a variable mass term, is deformed tending  to avoid
the regions with a large bare mass, and get an energetically
favorable position, concentrating inside or on the boundary of the
bag.

The bag conception should be applied for the Dirac wave function
on the KN background. Taking into account the discussed in sec.1.
peculiarities of the gravitating KN bag model, the
self-interacting Higgs field should be confined inside the bag. In
agreement with (\ref{Dir-sigma}), the vev of the Higgs field
$\sigma$ should give the mass term $m =  g \sigma $ to the Dirac
equation through the Yukawa coupling between the left-handed and
right-handed spinor field inside the bag, in full consistency with
the results of previous section. The corresponding Hamiltonian is
\be H(x) =\Psi^\dag (\frac 1{i} \vec \alpha \cdot \vec \nabla
+g\beta\sigma ) \Psi \label{Ham}, \ee and the energetically
favorable wave function has to be determined by minimization of
the averaged Hamiltonian ${\cal H} = \int d^3 x H(x) .$ Similar to
results of the SLAC-bag model, the one expected that the Dirac
wave function will be pushed from the region inside the bag, where
the bare mass $m =g\eta$ is large, towards a narrow zone at the
bag border.  Similarly to the case of the MIT and SLAC bags, the
narrow concentration of the Dirac wave function is admissible for
scalar potential, since it  does not lead to the Klein paradox.
Concrete form of the wave function will depend on the ratio of the
parameters $\sigma$ and $\eta$. In the strong coupling limit $g\to
\infty ,$ the wave function will concentrate on the shell of the
bag. The exact solutions of this kind are known only for
two-dimensional case, and the corresponding variational problem
for the KN soliton should apparently be solved numerically by
using the ansatz $\tilde \Psi =f(r)\Psi (x) ,$ where $f(r)$ is a
variable factor of the deformation, and  the Dirac bispinor $\Psi$
is formed by the Weyl spinors (\ref{phiY+}) and (\ref{phiY-})
aligned with two null congruences given by the Kerr theorem.

\section{String from deformation of a bag} Taking the bag
model conception, we should also accept the dynamical point of
view that the bags may easily be deformed \cite{SLAC,Giles}, and
deformations of the bag create stringy structures. The
deformations considered in the bag models are typically formations
of the bag into an open flux-tube string with radial and
rotational excitations. The known Dirac's model of an "extensible"
spherical electron \cite{DirBag} may be considered as a prototype
of the bag model. Under vanishing rotation, $a = 0 ,$ the KN
disk-like bag turns into the spherical Dirac "extensible" electron
model. The  non-rotating spherical KN bag has just the Dirac
radius $R$ corresponding to classical radius of the electron,
$R=r_e=e^2/2m .$ In fact, the disk-like bag of the KN rotating
source may be considered as a bag obtained by the rotational
stretch from the Dirac "extensible" spherical bag. Kerr's
parameter of rotation $a=J/m$ stretches the spherical bag to the
disk of the Compton radius $a= \hbar/2mc ,$ which indicates that
the KN bag should correspond to the zone of vacuum polarization of
a ``dressed'' electron. Since the degree of oblateness of the KN
bag turns out to be very close to $ \alpha= 137^{-1},$ the fine
structure constant acquires in the KN bag a geometrical
interpretation. Under stringy deformations the bag may acquire
oscillations similar to excitation of the strings, \cite{Giles}.
For the KN bag-like source, concentration of the wave function at
the border of the KN disk results in the appearance of the
circular light-like  string, similar to obtained by Sen
fundamental string to low energy heterotic string theory
\cite{BurSen}. \fn{The real and complex stringy structures of the
Kerr geometry were discussed in \cite{BurSPStr,BurDStr,BurTwi}.}
It may be shown that the lowest excitation of the Kerr closed
string creates a circulating singular point which may be
interpreted as a confined quark in the conception of the bag
models, or as a point-like bare electron with zitterbewegung of
the Dirac theory, either as an end point of an open circular
string (D0-brane \cite{BurDStr}) in the conception of string
theory.

\section{Conclusion}
Considering the problem of source of the KN solution we arrive at a
gravitating soliton model based on the Higgs model of broken symmetry,
which is similar to many other models of solitons, bags and Q-balls.
However, the requirement to retain the long-range KN gravitational
field outside the source enforces us to refuse from the usual quadratic
term of self-interaction and introduce especial supersymmetric scheme of
phase transition, in which symmetry is to be broken only inside the source
where it realizes a supersymmetric false-vacuum state. As a result, we
automatically obtain the flat space-time inside the source, avoiding
contradictions between gravity and the standard quantum theory.
The consequent treatment of the Dirac equation on the regularized KN background
exhibited three important peculiarities:

1) structure of the Dirac equation is close related with two-sheeted structure
of the Kerr geometry,

2) the Weyl components of the Dirac wave function are close related with twistor
structure of the Kerr geometry determined by the Kerr theorem,

3)  the Dirac equation acquires a variable mass term which find a strong theoretical
interpretation in the frame of bag models.

We conclude that the source of KN solution should be considered as a gravitating bag model,
and its further development should be based on theory of the bag models.

 The KN bag represents a gravitating extension of the Q-ball models, which were suggested in \cite{Kusen,VolkWohn,Grah} for electroweak sector of the standard model, and therefore, the gravitating KN bag can be considered as a step beyond the standard model towards its unification with gravity.

\medskip

\noindent \textbf{Comment added on December 20, 2014:}

 Jim Bogan informed me today that a geometrical model of the electron and its spin was published only just by the distinguished mathematician, Sir Michael Atiyah \& colleagues, http://arxiv.org/pdf/1412.5915.pdf
  but they take a different path than in my paper.
  I could note that their work is based on the remarkable properties of self-duality of the Taub-NUT solution and its connection with Dirac theory is defined by twistorial structure. All these properties  are present also in the Kerr-Newman solution, and the most part of their mathematical construction is to be related to the Kerr-Newman solution, too. Indeed, there is a remarkable common basis -- the Kerr-NUT solution which lies beyond of the both lines of consideration, and we consider ideed different parts of this basic solution.
   The Kerr-Newman solution is considered in this paper in the Kerr-Schild form of metric which has an
  auxiliary Minkowski  background, allowing us to present geometry of the extended electron in the real physical  space-time. The  NUT-part is apparently important, but it cannot be represented in the Kerr-Schild form, and so far it drops out from my consideration. On the other hand, the treatment of Sir Michael Atiyah et al is based on the NUT-part of the Kerr-NUT solution which requires a more abstract mathematical treatment and turns out to be far from our physical space-time. In my opinion the spin of the electron is related basically with the Kerr-Newman geometry, while the NUT-part may be important for self-interaction. I am thankful to Jim Bogdan for paying my attention to this paper.

 \subsection*{Acknowledgments} This
work is supported by the RFBR grant 13-01-00602. Author thanks
 Theo M. Nieuwenhuizen, Yuri Rybakov and Bernard Whiting for interest to this work and useful conversations.


\small
\begin{thebibliography}{99}
\bibitem{BHpart}
 G. 't Hooft,  The black hole interpretation of string theory, \emph{Nucl. Phys.} \textbf{B 335}, 138 (1990);
 C.F.E. Holzhey and F. Wilczek, Black Holes as Elementary Particles, \emph{Nucl. Phys.} \textbf{B 380}, 447
 (1992); A. Sen, Extremal black holes and elementary string states, \emph{Mod. Phys. Lett.} \textbf{A 10} 2081 (1995).


\bibitem{Car} B. Carter,  Global structure of the
Kerr family of gravitational fields {\it
  Phys.\ Rev.} {\bf 174} 1559 (1968).

\bibitem{Keres} H. Keres, To physical interpretation of the
solutions to Einstein equations {\it Zh.Exp. i Teor.Fiz (ZhETP)}
{\bf 52} 768 (in Russian) (1967).

\bibitem{Isr}
W. Israel, Source of the Kerr metric {\it  Phys.\ Rev.}  D {\bf 2}
641 (1970).

\bibitem{Ham} V. Hamity, An interior of the Kerr metric,
{\it Phys. Lett.} A {\bf 56}, 77, (1976).


\bibitem{Lop} C. A. L\'opez (1984) An extended model of the electron in general relativity
  {\it Phys.\ Rev.}  D {\bf 30} 313 (1984).



\bibitem{BurSol} A. Burinskii,
Regularized Kerr-Newman Solution as a Gravitating Soliton,
\emph{J. Phys. A: Math. Theor.} {\bf 43} (2010) 392001 [arXiv:
1003.2928].



\bibitem{GRos}  G. Rosen, Particlelike Solutions to Nonlinear Complex Scalar Field Theories
with Positive-Definite Energy Densities. \emph{J. of Math. Phys.}
\textbf{9} (7) 996 (1968), doi:10.1063/1.1664693 .


\bibitem{Coleman} S. Coleman, Q-Balls, \emph{Nuclear Physics} \textbf{B 262} (2) 263 (1985),


\bibitem{Kusen} A. Kusenko, Solitons in the supersymmetric extensions of the standard
model, \emph{Phys.Lett.} \textbf{B405} 108 (1997).



\bibitem{VolkWohn} M.Volkov and E. W\"ohnert, Spinning Q-balls, \emph{ Phys.Rev.} D {\bf 66}, 085003
(2002).

\bibitem{Grah} N.Graham, An Electroweak Oscillon, \emph{Phys.Rev.Lett} {\bf 98}, 101801 (2007).

\bibitem{BurMag} A. Burinskii and G. Magli, Kerr-Schild approach to the boosted Kerr solutions, \emph{Phys. Rev.} \textbf{D 61}  044017 (2000).


\bibitem{BurTMP} A. Burinskii, Stringlike structures in Kerr-Schild geometry: The N=2 string, twistors, and Calabi-Yau twofold,
\emph{Theor. Math. Phys.}, \textbf{177}(2), 1492 - 1504, (2013).

\bibitem{DirRad} P. A. M. Dirac, \emph{Proc. R. Soc. London,} Ser.\textbf{ A 167}, 148
 (1938).

\bibitem{FeynST} R. Feynman, Space-Time Approach to Quantum Electrodynamics,
\emph{Phys. Rev. }   \textbf{76} 769 (1949).

\bibitem{MTW} Ch.W. Misner, K.S. Thorne and J.A. Wheeler, \emph{Gravitation} v.3,
 San Francisco: W. Freeman, 1973.

\bibitem{BurA} A. Burinskii, First Award of Gravity Research
Foundation, 2009,  \emph{Gen. Rel. Gravit.} {\bf 41} 2281 (2009), arXiv:
0903.3162[gr-qc]

\bibitem{DKS} G. C. Debney, R. P. Kerr and A. Schild
  Solutions of the Einstein and Einstein-Maxwell equations {\it
  J.\ Math.\ Phys.}  {\bf 10} 1842 (1969).

\bibitem{KraSte}
D.Kramer, H.Stephani, E. Herlt, M.MacCallum, \emph{Exact Solutions
of Einstein's Field Equations,} Cambridge Univ. Press, Cambridge
1980.

\bibitem{PenRin} R. Penrose, Twistor Algebra, {\it J. Math. Phys.} {\bf 8} 345 (1967); R.
Penrose and W. Rindler, \emph{Spinors and Space-time, Vol. 2:
Spinor and twistor methods in space-time geometry,} Cambridge
University Press, Cambridge U.K. (1986), pg. 501.

\bibitem{WesBag} J. Wess, J. Bagger,  {\it Supersymmetry and
Supergravity} (Princeton Univ. Press, Princeton, New Jersey),
1983.

\bibitem{Mor} J. R. Morris, \emph{Phys. Rev.} \textbf{D 53} 2078 (1996) [arXiv:hep-ph/9511293].

\bibitem{EinFinkel} S. Einstein and R. Finkelstein, Lorentz
covariance and the Kerr-Newman geometry, \emph{Phys. Rev.}
\textbf{D 15} 2721 (1977).

\bibitem{BLP}
V.B. Berestetsky, E.M. Lifshitz, L.P. Pitaevsky, \emph{Quantum
Electrodynamics ( Course Of Theoretical  Physics, 4)}, Oxford, Uk:
Pergamon ( 1982).

\bibitem{MIT} A. Chodos, R. L. Jaffe, K. Johnson, C. B. Thorn, and V. F. Weisskopf,
New extended model of hadrons, \emph{Phys. Rev.} \textbf{D 9},
3471 (1974).

\bibitem{SLAC} W. A. Bardeen, M. S. Chanowitz, S. D. Drell, M. Weinstein, and T.-M. Yang, Heavy quarks and strong binding: A field theory of hadron structure, \emph{Phys. Rev.} \textbf{D 11}, 1094 (1974).

\bibitem{Giles} R.C. Giles, Semiclassical dynamics of the ``SLAC'' bag,
 {\it Phys.\ Rev.\  D} {\bf 13} (1976) 1670,

\bibitem{DirBag} P.A.M. Dirac, An Extensible Model of the Electron, \emph{Proc. R.
Soc. Lond.} \textbf{A 268},  57-67 (1962).

\bibitem{BurSen} A.~Burinskii, Some properties of the Kerr solution to low-energy string theory,
  {\it Phys.\ Rev.\  D} {\bf 52} (1995) 5826, [arXiv:hep-th/9504139].

\bibitem{BurSPStr} A.Ya. Burinskii,
Kerr Spinning Particle, Strings and Superparticle Models.
\emph{Phys. Rev.}{\bf D 57} (1998) 2392.

\bibitem{BurDStr} A.~Burinskii,
Orientifold D-String in the Source of the Kerr Spinning Particle
{\it Phys.\ Rev. } {\bf D 68} (2003) 105004 [arXiv:hep-th/0308096].

\bibitem{BurTwi} A.~Burinskii,
Twistorial analyticity and three stringy systems of the Kerr spinning particle,
{\it Phys.\ Rev. } {\bf D 70} (2004) 086006,
[arXiv:hep-th/0406063].

\end{thebibliography}
\end{document}